\documentstyle[12pt,epsfig,amsfonts,here]{article}

\def\Journal#1#2#3#4{{#1} {\bf #2}, #3 (#4)}

\def\NCA{{\em Nuovo Cimento} A}

\def\PLB{{\em Phys. Lett.}  B}

\def\PRD{{\em Phys. Rev.} D}
\def\ZPC{{\em Z. Phys.} C}
\def\EPG{{\em Eur.Phys.J.} C}

\def\be{\begin{equation}}
\def\ee{\end{equation}}
\def\bea{\begin{eqnarray}}
\def\eea{\end{eqnarray}}

\begin{document}
\begin{center}
{\large\bf COULOMB INTERFERENCE IN HIGH-ENERGY $pp$ AND $\bar p p$ SCATTERING.}
\vskip 1.cm
{V. A. PETROV$^{a,}$\footnote{\it E-mail: petrov@mx.ihep.su},
 E. PREDAZZI$^{b,}$\footnote{\it E-mail: predazzi@to.infn.it}
 and A. PROKUDIN$^{a,b,}$\footnote{\it E-mail: prokudin@to.infn.it}}
\vskip 0.5cm
{\small\it
(a) Institute For High Energy Physics,\\
142281 Protvino,  RUSSIA}
\vskip 0.2cm

{\small\it
\vskip 0.2cm
(b)  Dipartimento di Fisica Teorica,\\
Universit\`a Degli Studi Di Torino, \\
Via Pietro Giuria 1,
10125 Torino, \\
ITALY\\
and\\
Sezione INFN di Torino,\\
 ITALY\\}
\vskip 0.5cm


\parbox[t]{12.cm}{\footnotesize An analysis of the Coulombic amplitude
and its interference with the nuclear amplitude which is driven
by the three-component Pomeron \cite{threepomerons}
is presented. It is shown that different approaches 
towards the Coulomb phase evaluation give approximately uniform result at all
energies and the difference is negligible at {\bf RHIC} and
{\bf LHC} energies. We show that the use of the amplitude which was fitted to accomodate
nucleon data only (in the region
$0.01 \le |t| \le 14.5 \; (GeV^2)$) combined with the Coulomb amplitude, reproduces the existing 
data in the Coulomb interference domain quite accurately without any adjustement 
of the parameters. As a consequence, we predict the differential
cross section in the region of the Coulomb nucleon interference for both 
{\bf RHIC} and {\bf LHC} energies.
}
\end{center}

\section{INTRODUCTION}
In Ref. \cite{threepomerons} an eikonal model of a three-component Pomeron
has been suggested and successfully used for describing 
the high energy $pp$ nad $\bar pp$  data in the region 
of large momentum transfer
$0.01 \le |t| \le 14.5 \; (GeV^2)$ . In this paper we apply the model
to the region of small momentum transfer $0 \le |t| \le 0.01 \; (GeV^2)$.

The problem is a proper account of the Coulomb interaction which is most
 important at the smallest $|t|$. The standard way to do this is
to represent the whole scattering amplitude $T(s,t)$ which is dominated by the Coulomb force at
low momentum transfer and by the hadronic force at higher momentum transfer
as

\be
T(s,t)=T^N(s,t)+e^{i\alpha\Phi}T^C(s,t),
\ee
where if we normalize the scattering amplitude so that
\be
\displaystyle \frac{d\sigma}{dt} = \frac{\vert T(s,t) \vert ^2}{16\pi s^2},
\ee
the Born Coulomb amplitude for $pp$ and $\bar pp$  scattering is
\be
\displaystyle T^C(s,t) = \mp\frac{8\pi \alpha s}{|t|}.
\ee
The upper (lower) sign corresponds
to the scattering of particles with the same (opposite) charges. $T^N(s,t)$ stands for purely strong interaction amplitude, and
the phase $\Phi$ depends generally on energy, 
the momentum trasfer and on the properties 
of $T^N$. The study of the Coulomb nuclear interference is very important
for extracting the real part of the strong interaction amplitude.
    
The issue of Coulombic amplitude and its interference with the nuclear component
was addressed in many papers in the past. West and Yennie~\cite{West} examined
the Coulomb-nuclear interference using Feynman diagrams. R. Cahn~\cite{Cahn}
considered the same task in the framework of the Eikonal model; his
results were quite convincing (though the modifications were very small
compared to ~\cite{West}) and proved that the Eikonal model is a very convenient
basis for analyzing the Coulomb-nuclear interference.

The Coulombic phase has atracted the attention of many authors. Using the 
WKB approach in potential theory, Bethe~\cite{Bethe} derived, for 
proton-nucleus scattering, the expression
\be
\Phi = 2 \ln (1.06/|{\bf k_1}| b \Theta),
\ee
where $|{\bf k_1}|$ is the c.m. momentum, $b$ is the range of strong-interaction
forces defined by the size of the nucleus, and $\Theta$ the c.m. scattering angle. 
Similar results were derived within potential theory by several authors~\cite{Rix1}, 
~\cite{Islam}.

A relativistic derivation of the phase was attempted by Soloviev~\cite{Soloviev}, 
who obtained
\be
\Phi = 2 \ln (2/\Theta),
\ee
a result which differs considerably from the result of Bethe~\cite{Bethe}. 
Utilizing the same technique, Rix and Thaler~\cite{Rix} derived a result quantitevely 
close to that of Bethe~\cite{Bethe}.

West and Yennie~\cite{West}, as already mentioned, obtained the phase of
Coulomb-nuclear interference using Feynman diagrams. For a conventional 
parametrization $T^N(s,t)\sim exp(-B|t|/2)$, the result of West and 
Yennie~\cite{West} reads
\be
\Phi_{W-Y}=\mp \Big[\ln (B|t|/2)+\gamma+O(B|t|)\Big],
\label{eq:westphase}
\ee
where $\gamma=0.577...$ is Euler's constant. The upper (lower) sign
corresponds  to the scattering of $pp$ ($\bar p p$).

Cahn~\cite{Cahn} analized also the effect of the electromagnetic form factor and
obtained a general expression for the phase. The results of~\cite{Cahn} were in
complete agreement with~\cite{West} and especially the formula 
(\ref{eq:westphase}) was derived from a rather different perspective.

The main difference from the result~\cite{West} is a shift of the Coulomb 
amplitude due to the form factor's influence on the phase. If we introduce the 
electromagnetic form factor of the proton, the Born term of the Coulomb 
amplitude 
has the following form
\be
\displaystyle T^C(s,t) = \mp \frac{8\pi \alpha s}{|t|}f^2(|t|),
\ee
where the form factor may be chosen as
\be
f(|t|) = e^{-2|t|/\Lambda^2}, \;\; \Lambda^2=0.71\;GeV^2.
\ee
In this case the Coulomb phase has the following form
\bea
\nonumber
\Phi_{Cahn}=\mp \Big[\ln \Big(\frac{B|t|}{2}\Big)+\gamma+
\ln\Big(1+\frac{8}{B\Lambda^2}\Big) \\ 
+(4|t|/\Lambda^2)\ln(4|t|/\Lambda^2)+2|t|/\Lambda^2\Big]. 
\label{eq:cahnphase}
\eea

All these results were obtained under the assumption that $|t|\rightarrow 0$.
The derivation of the phase in a large domain of momentum transfer was attempted
by Selyugin~\cite{Selyugin}, and Kopeliovich and Tarasov~\cite{Kop}. In the 
region of interest for the present paper $0\le|t|\le 0.01 \;GeV^2$, the latter 
results are similar to that of Cahn~\cite{Cahn} and the main difference is 
in the dip region of the differential cross-sections.
The phase obtained in \cite{Selyugin} 
accurately takes into account the dipole form factor 
\be
f(|t|) = \Big[\frac{\Lambda^2}{\Lambda^2+|t|}\Big]^2, \;\; \Lambda^2=0.71\;GeV^2,
\ee
and the complicated structure of the nucleon amplitude.

The phase has the following form
\bea
\nonumber
\Phi_{Selyugin}=\mp \Big[\ln\frac{|t|}{4}+2\gamma-\nu_s  
-i\frac{8\pi s}{T_N(s,t)}\int_0^\infty bdb J_0(b\sqrt{-t})\delta_C(s,b)\cdot 
\\ \nonumber
\Big(e^{2i\delta_N(s,b)}-1\Big)\Big], 
\label{eq:selphase}
\eea
where 
\be
\nu_s = 0.11 \log(1+400|t|)
\ee
takes into account influence of the form factor on the pure coulombic
phase, and
\be
\delta_C(s,b) = \ln b+ K_0(b\Lambda)+\frac{11}{12}b\Lambda K_1(b\Lambda)+
\frac{5}{24}b^2\Lambda^2 K_0(b\Lambda)+ \frac{1}{48}b^3\Lambda^3 K_1(b\Lambda).
\ee

The prescription of West and Yennie~\cite{West} was successfully
used by several authors \cite{Jenya} and 
\cite{Wu} for describing differential cross sections
in low $t$ region.

In what follows we will investigate four different cases of the Coulomb
phase -- the phase calculated with the nucleon amplitude of 
the model \cite{threepomerons} (which does not acquire any additional parameter) with the 
prescription of West and Yennie~\cite{West},  the phase calculated with 
prescription of Cahn~\cite{Cahn}, the prescription of Selyugin \cite{Selyugin},
and the phase equal to zero.

\section{THE NUCLEAR AMPLITUDE}

We believe that any nuclear amplitude that is capable of a high accuracy 
description of the combined set of high energy $pp$ and $\bar p p$ data (total 
and differential cross sections, $\rho$ parameter etc.) over the entire $|t|$ 
spectrum, if properly combined with the correct Coulomb amplitude {\it must}
account well for the data in the interference region. That this is so, we will
prove using the particular nuclear amplitude which has been derived in 
~\cite{threepomerons} to describe total and differential cross sections at 
high energies ($\sqrt s \ge 10 GeV$) in the range of momentum transfer 
$0.01<|t|<14.5\;GeV^2$ using the eikonal approach (another one could have been
the amplitude of ~\cite{DGMP}).
We just write the nuclear amplitude of ~\cite{threepomerons}  
\begin{equation}
T(s,\vec b)=\frac{e^{2i\delta (s,\vec b)}-1}{2i}\; ,
\label{eq:ampl}
\end{equation}
where the eikonal has the following form
\be
\delta_{pp}^{\bar p p}(s,b) = \delta^+_{{\Bbb P}_1}(s,b)+
\delta^+_{{\Bbb P}_2}(s,b)+
\delta^+_{{\Bbb P}_3}(s,b)
\mp \delta^-_{\Bbb
O}(s,b)+\delta^+_{
f}(s,b)\mp \delta^-_{\omega}(s,b).
\label{eq:modeleik}
\ee
We refer the reader to the original literature for details; let us simply 
recall that here $\delta^+_{{\Bbb P}_{1,2,3}}(s,b)$ are Pomeron contributions. 
The superscript `$+$' denotes C even trajectories (the Pomeron trajectories have 
quantum numbers $0^+J^{++}$), while `$-$' denotes  C odd trajectories. 
$\delta^-_{\Bbb O}(s,b)$ is the Odderon contribution (i.e. the C odd partner of
the Pomeron whose quantum numbers are $0^-J^{--}$); $\delta^+_{ f}$ and 
$\delta^-_{\omega}(s,b)$ are the contributions of secondary Reggeons, ($f$
as representative of the $C=+1$ families and $\omega$ of the $C=-1$).

In order to relate $t$- and $b$-spaces one proceeds via Fourier-Bessel
transforms
\be
\begin{array}{r}
\displaystyle \hat f(t)= 4 \pi s\int_{0}^{\infty} db^2 J_0(b\sqrt{-t}) f(b)\; , \\
\\
\displaystyle f(b)= \frac{1}{16 \pi s}\int_{-\infty}^{0} dt J_0(b\sqrt{-t}) \hat f(t) \; .
\end{array}
\label{eq:fb}
\ee

Using this parametrisation we obtain the following expressions 
for the Coulomb phase
\be
\displaystyle \Phi_{W-Y}=\mp \frac{\sum_i \delta_i(s,t=0)\Big[
\ln \Big(\frac{\rho_i(s)|t|}{4}\Big)+\gamma\Big]}{\sum_i \delta_i(s,t=0)},
\label{eq:ourwest}
\ee
and
\bea
\nonumber
\displaystyle \Phi_{Cahn}=\mp \frac{\sum_i \delta_i(s,t=0)\Big[
\ln \Big(\frac{\rho_i(s)|t|}{4}\Big)+\gamma+\ln\Big(1+\frac{16}{\rho_i(s)\Lambda^2}\Big)\Big]}{\sum_i \delta_i(s,t=0)}\\ 
\mp(4|t|/\Lambda^2)\ln(4|t|/\Lambda^2)\mp 2|t|/\Lambda^2 ,
\label{eq:ourcahn}
\eea
where $\rho_i^2 = 4\alpha_i'(0) \ln s/s_0+r_i^2$, $i=P_1,P_2,P_3, O, f,\omega$.
The upper (lower) signs correspond  to $pp$ ($\bar p p$).

Crossing symmetry is restored by replacing
$s \rightarrow s e^{-i\pi/2}$. We introduce the dimensionless variable
\be
\tilde s = \frac{s}{s_0}e^{-i\frac{\pi}{2}}\; ,
\ee
in terms of which we give each $C+$ and $C-$ contribution
with its appropriate signature factor in the form

\bea
\displaystyle \delta^+ (s,b)=i\frac{c}{s_0}
\tilde s^{\alpha(0)-1}\frac{e^{-\frac{b^2}{\rho^2}}}{4\pi \rho^2}\; ,
\label{eq:eikonalform} \\
\nonumber
\rho^2 = 4\alpha'(0) \ln\tilde s+r^2\; , \\
\nonumber
(C = +1)\; ;
\\
\displaystyle \delta^- (s,b)=\frac{c}{s_0}
\tilde s^{\alpha(0)-1}\frac{e^{-\frac{b^2}{\rho^2}}}{4\pi \rho^2}\; ,
\label{eq:eikonalform1}
\\
\nonumber
\rho^2 = 4\alpha'(0) \ln\tilde s+r^2\; , \\
\nonumber
(C = -1)\; .
\eea

\section{RESULTS}

In \cite{threepomerons}, the adjustable parameters have been fitted 
over a set of 982 $pp$ and $\bar p p$ data\footnote{The data are available at \\ 
REACTION DATA Database {\it http://durpdg.dur.ac.uk/hepdata/reac.html} \\
CROSS SECTIONS PPDS database {\it http://wwwppds.ihep.su:8001/c5-5A.html}\\
{\it http://pdg.lbl.gov/2000/contents$_{-}$plots.html}} of both forward observables 
(total cross-sections $\sigma_{tot}$, and $\rho$ -- ratios of real to 
imaginary part of the amplitude) in the range $8.\le\sqrt{s}\le 1800.\; GeV$ 
and angular distributions
($\frac{d\sigma}{dt}$) in the ranges $23.\le\sqrt{s}\le 1800.\; GeV$,
$0.01\le |t|\le 14.\; GeV^2$. A good $\chi^2/d.o.f. =2.60$ was obtained and
the parameters are given in Table~\ref{tab:1}. 

{
\begin{table}[H]
\begin{center}
\begin{tabular}{|l|l|l|l|}
\hline
& {\bf Pomeron$_{\bf 1}$} & & {\bf $\bf f$-Reggeon}  \\
\hline
$\Delta_{{\Bbb P}_1}$ & $0.0578\pm0.0020$ & $\Delta_{f}$ & $-0.31$ (FIXED) \\
$c_{{\Bbb P}_1}$ & $53.007\pm0.795$  & $c_{f}$ & $191.69\pm2.12$            \\
$\alpha'_{{\Bbb P}_1}$& $0.5596\pm0.0078\;(GeV^{-2})$  & $\alpha'_{f}$& $0.84\;(GeV^{-2})$ (FIXED)    \\
$r^2_{{\Bbb P}_1}$& $6.3096\pm0.2522\;(GeV^{-2})$&$r^2_{f}$ & $31.593\pm1.099\;(GeV^{-2})$  \\
\hline
& {\bf Pomeron$_{\bf 2}$}& & {\bf $\bf \omega$-Reggeon}  \\
\hline
$\Delta_{{\Bbb P}_2}$ & $0.1669\pm0.0012$  & $\Delta_{\omega}$ & $-0.53$ (FIXED)   \\
$c_{{\Bbb P}_2}$ & $  9.6762\pm0.1600$    & $c_{\omega}$ & $-174.18\pm2.72$           \\
$\alpha'_{{\Bbb P}_2}$& $0.2733\pm0.0056\;(GeV^{-2})$    & $\alpha'_{\omega}$& $0.93\;(GeV^{-2})$ (FIXED)     \\
$r^2_{{\Bbb P}_2}$& $3.1097\pm0.1817\;(GeV^{-2})$ &$r^2_{\omega}$ & $7.467\pm1.083\;(GeV^{-2})$  \\
\hline
& {\bf Pomeron$_{\bf 3}$}& & \\
\hline
$\Delta_{{\Bbb P}_3}$ & $0.2032\pm0.0041$  &$s_0$& $1.0\;(GeV^2)$ (FIXED)   \\
$c_{{\Bbb P}_3}$ & $1.6654\pm0.0669$       & &       \\
$\alpha'_{{\Bbb P}_3}$& $0.0937\pm0.0029\;(GeV^{-2})$   & &     \\
$r^2_{{\Bbb P}_3}$& $2.4771\pm0.0964\;(GeV^{-2})$ & & \\
\hline
& {\bf Odderon}& & \\
\hline
$\Delta_{{\Bbb O}}$ & $0.19200\pm0.0025$  & &  \\
$c_{{\Bbb O}}$ & $0.0166\pm0.0022$      & &        \\
$\alpha'_{{\Bbb O}}$& $0.048\pm0.0027\;(GeV^{-2})$   & &     \\
$r^2_{{\Bbb O}}$& $0.1398\pm0.0570\;(GeV^{-2})$ & & \\
\hline
\end{tabular}
\end{center}
\caption{Parameters obtained in \cite{threepomerons}. 
\label{tab:1}}
\end{table}}

We now consider the complete set of data
 including the Coulomb region 
which consists of 2158 data\footnote{The data are available at \\ 
REACTION DATA Database {\it http://durpdg.dur.ac.uk/hepdata/reac.html} \\
CROSS SECTIONS PPDS database {\it http://wwwppds.ihep.su:8001/c5-5A.html}\\
{\it http://pdg.lbl.gov/2000/contents$_{-}$plots.html}}. To these data, we apply the model including the 
Coulomb part with its phase and we simply plot the physical quantities
without any additional fitting.

\begin{figure}[H]
\vskip -1.cm
\parbox[c]{6.5cm}{\vspace*{ -0.5cm} \epsfxsize=65mm \epsffile{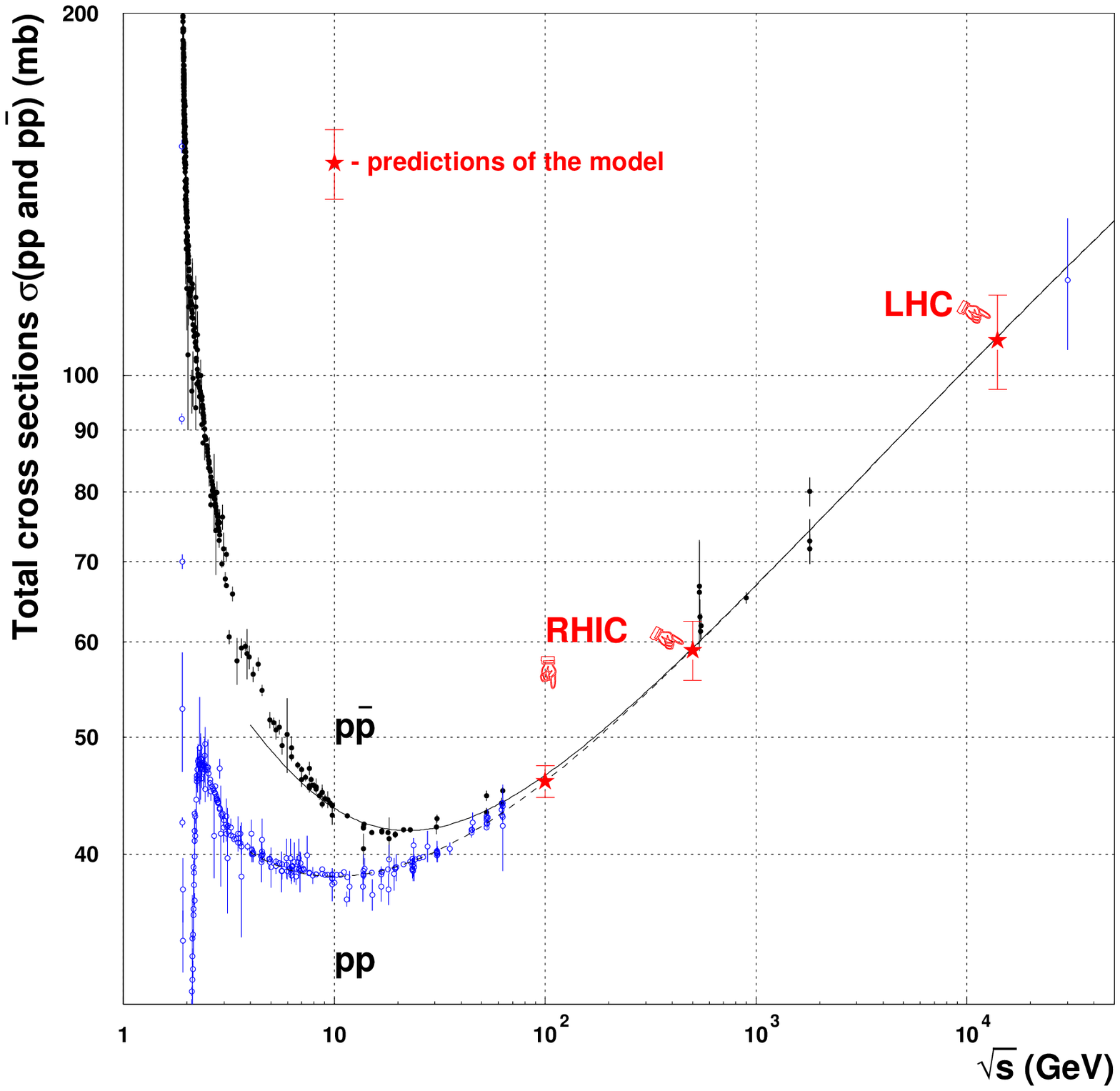}}
\hfill~\parbox[c]{6.5cm}{\vspace*{ -0.5cm} \epsfxsize=65mm 
\epsffile{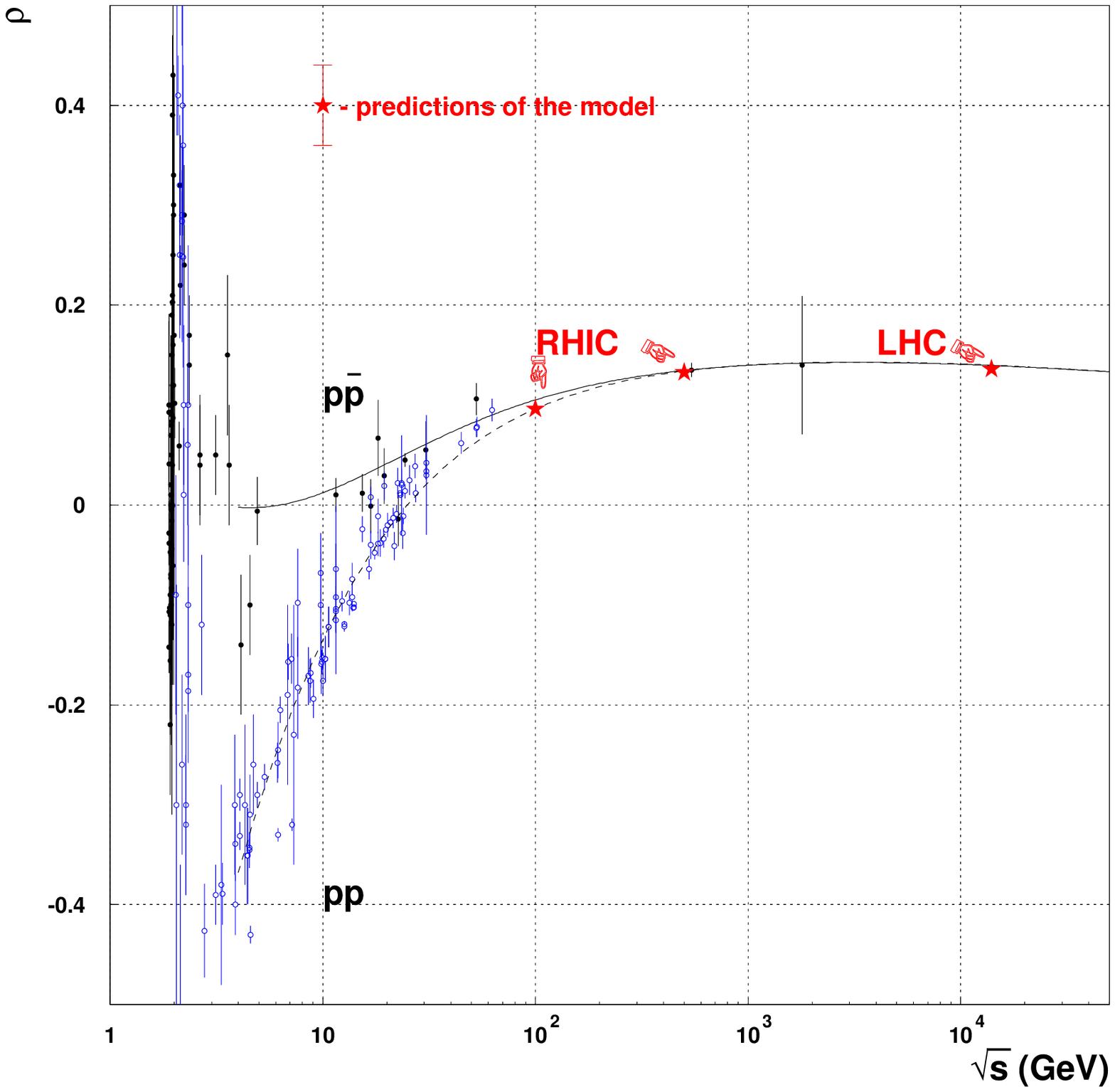}}
\vskip -1.5cm
\parbox[t]{6.5cm}{\caption{Total cross sections of $pp$ scattering (hollow circles)  
and $\bar p p$ scattering (full circles) and corresponding curves 
in the model of Ref.\cite{threepomerons}.
\label{fig:tot}
}}
\hfill~\parbox[t]{6.5cm}{\caption{Ratios of the real to the imaginary part of the forward $pp$ 
scattering amplitude (hollow circles)  and $\bar p p$ scattering  amplitude 
(full circles) and corresponding curves in the model of Ref.
\cite{threepomerons}.
\label{fig:reim}}}
\end{figure}

The total cross sections and the ratios of real to imaginary parts of the 
forward amplitude are presented in Figs. \ref{fig:tot} and \ref{fig:reim}.


In order to compare different approaches to the Coulomb phase, we have
calculated the $\chi^2$ for the region of low 
$|t|:\;\; 0\leq |t|\leq 0.01\; (GeV^2)$ in four different cases:
\begin{enumerate}
\item The Coulomb phase is equal to zero.
\item The Coulomb phase is calculated with the prescription of
West and Yennie (\ref{eq:ourwest}).
\item The Coulomb phase is calculated with the prescription of
Cahn (\ref{eq:ourcahn}).
\item The Coulomb phase is calculated with the prescription of
Selyugin (\ref{eq:selphase}).
\end{enumerate}

The results may be found in Table~\ref{tab:2}.
\begin{table}[H]
\begin{center}
\begin{tabular}{|l|l|l|}
\hline
Coulomb phase & Number of points & $\chi^2$ per point \\
\hline
$\Phi=0$ & $604$ & $3.49$ \\
$\Phi_{W-Y}$ (\ref{eq:ourwest}) & $604$ & $2.09$ \\
$\Phi_{Cahn}$(\ref{eq:ourcahn}) & $604$ & $1.86$ \\
$\Phi_{Selyugin}$(\ref{eq:selphase}) & $604$ & $1.84$ \\
\hline
\end{tabular}
\end{center}
\caption{ $\chi^2$ per point for the  region of
low $|t|:\;\; 0\leq |t|\leq 0.01\; (GeV^2)$. \label{tab:2}}
\end{table}

\begin{figure}[H]
\vskip -1.5cm
\parbox[c]{6.5cm}{\vspace*{ -0.5cm} \epsfxsize=65mm \epsffile{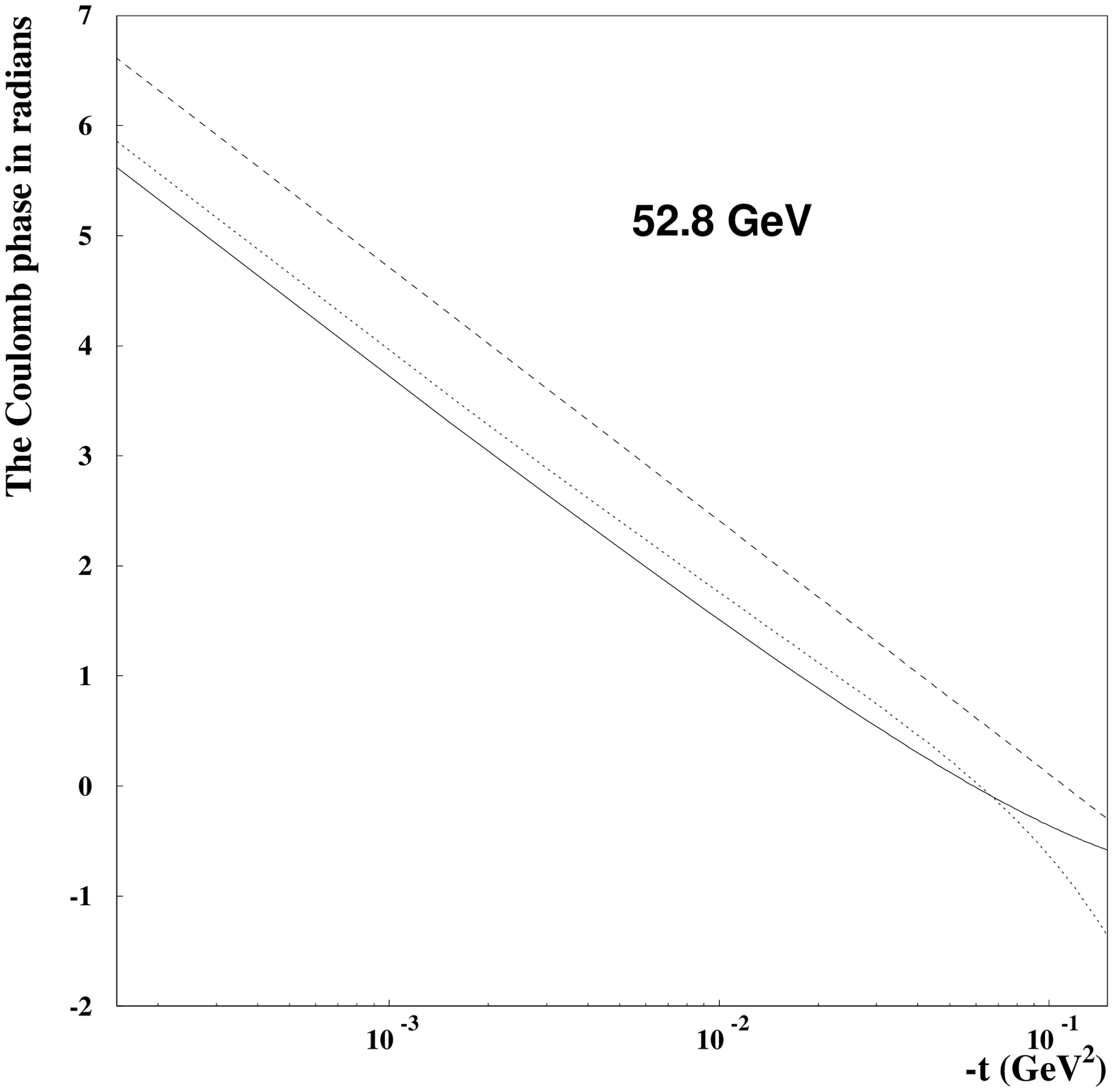}}
\hfill~\parbox[c]{6.5cm}{\vspace*{ -0.5cm} \epsfxsize=65mm 
\epsffile{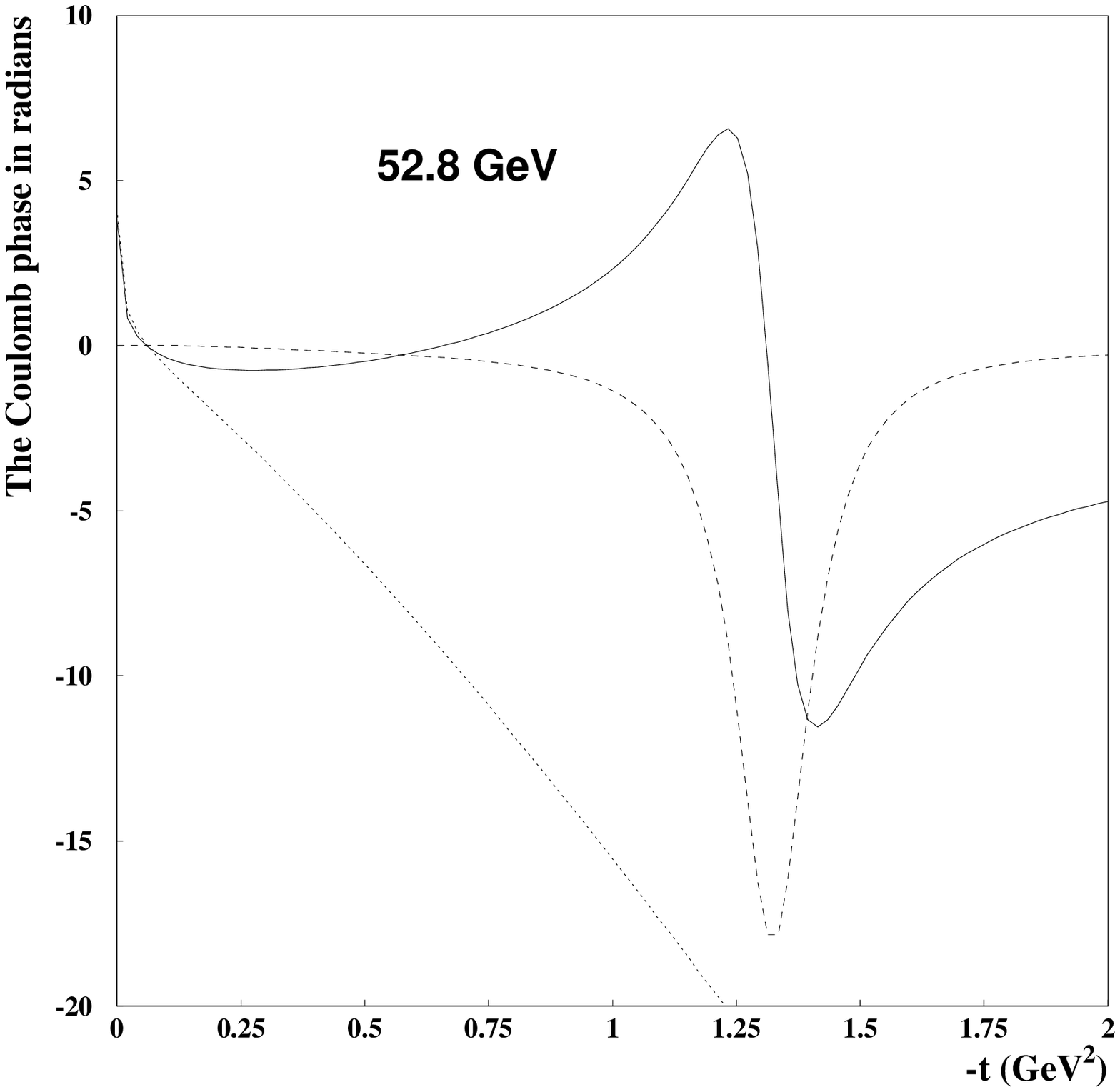}}
\vskip -1.7cm
\parbox[t]{6.5cm}{\caption{\footnotesize The Coulomb phase calculated in the framework
of Selyugin \cite{Selyugin} (the solid line) and of Cahn \cite{Cahn}
(the dotted line) and of West and Yennie \cite{West}
(the dashed line) in the region of small $t$. $\sqrt{s}=52.8\; GeV$
\label{fig:phase}}}
\hfill~\parbox[t]{6.5cm}{\caption{\footnotesize The Coulomb phase calculated in the framework
of Selyugin \cite{Selyugin} , real part (the solid line) and imaginary part
(the dashed line) and in that of Cahn \cite{Cahn}
(the dotted line) in the region of high $t$.  \label{fig:phase2}}}
\end{figure}

As is seen from the Table~\ref{tab:2}, the experimental data marginally 
``prefer'' the Coulomb phase calculated with the prescription of 
Cahn \cite{Cahn} and Selyugin \cite{Selyugin} over that of West and Yennie but taking the Coulomb phase 
equal to zero is excluded by the data and this is gratifying on physical 
grounds. The difference between the phases of \cite{Cahn} and  \cite{Selyugin}
is negligible in the small $|t|$ region of the interest (as may be seen in Fig.
\ref{fig:phase}), though the exact result
obtained in \cite{Selyugin} over the whole kinematical
region of $|t|$ shows that the difference is drastic in the region of 
the diffractive
dip (as seen in Fig. \ref{fig:phase2}).

Apart from this, the Coulomb phase calculated in the Selyugin approach 
exhibits a non trivial behaviour of the real part (which has a zero
in the point of the diffractive dip), and of the imaginary part (which has a minimun
in the same point) Fig. \ref{fig:phase2}. 

The diffrence between the approaches of Cahn \cite{Cahn} 
and of West and Yennie \cite{West} is more pronounced even in the
region of low $t$ as may be seen in Fig. \ref{fig:phase}.

It is of some interest to perform a more detailed comparison of the $\chi^2$
 derived  for the region of low momemtun transfer. This can be found in 
Table~\ref{tab:3} for $\Phi_{W-Y}$, $\Phi_{Cahn}$,
and $\Phi_{Selyugin}$.
 
With all due caution, a general pattern emerges; some data points have 
anomalously large $\chi^2$ values and these are the same in all cases. 

\begin{figure}[H]
\vskip -1.5cm
\parbox[c]{6.5cm}{\vspace*{ -0.5cm} \epsfxsize=65mm \epsffile{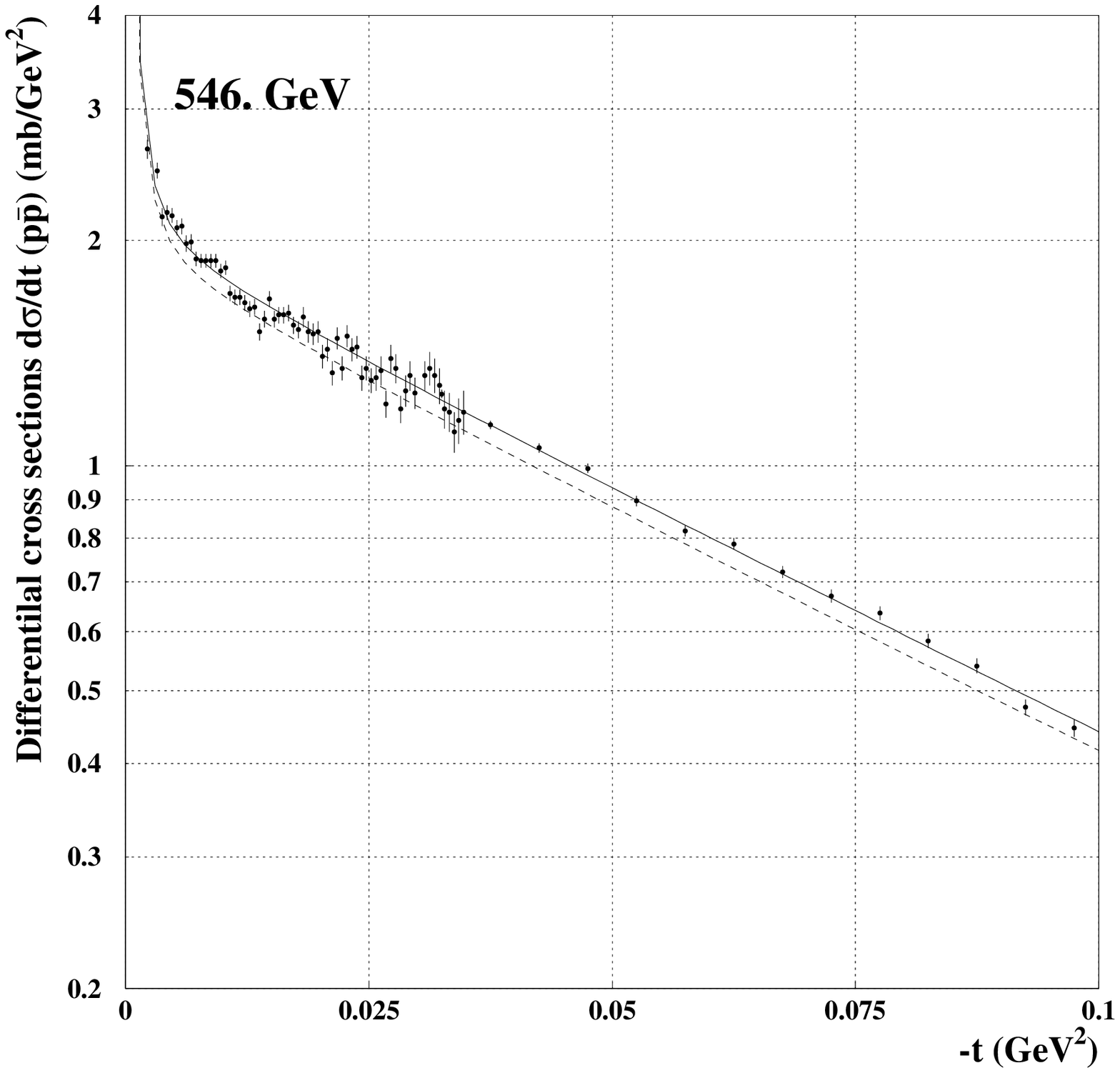}}
\hfill~\parbox[c]{6.5cm}{\vspace*{ -0.5cm} \epsfxsize=65mm 
\epsffile{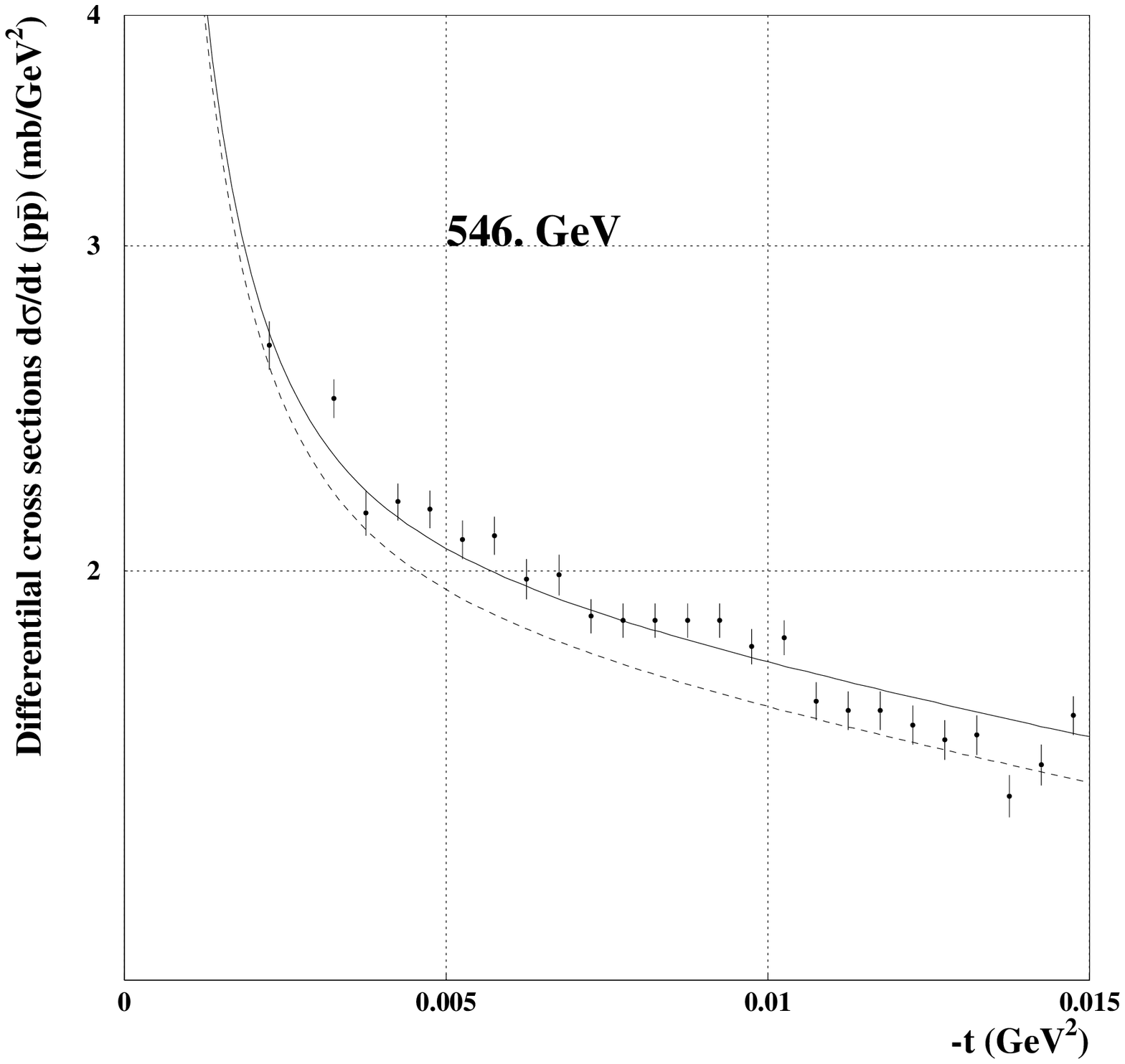}}
\vskip -1.7cm
\parbox[t]{6.5cm}{\caption{\footnotesize Differential cross section of $\bar p p$ scattering
and curves corresponding to its description in the model.
The dashed line represents our predictions. The solid curve is our predictions
normalized by a factor 1.06. 
\label{fig:pbarp546}}}
\hfill~\parbox[t]{6.5cm}{\caption{\footnotesize The same as
in Fig. \ref{fig:pbarp546} for low $t$.  \label{fig:pbarp1546}}}
\end{figure}

The $\bar p p$ differential cross section at $\sqrt{s}= 546 \; GeV$, for instance, sticks out conspicuously. The explanation is the normalisation of
the cross section. The sysytematical error allows to change the normalisation
of the data in the $\pm 10\% $ corridor. If we could 
multiply our predictions by
a factor 1.06, we would have a good description of this set of data,
as is seen in Fig. \ref{fig:pbarp546} and \ref{fig:pbarp1546}.



{\footnotesize
\begin{table}[H]
\begin{center}
\begin{tabular}{|l|l|l|l|l|l|l|}
\hline
&{\bf Data} & $\sqrt{s}\:(GeV)$ & $0\leq |t|\leq 0.01\:(GeV^2)$  &
{\bf $\chi^2/{\rm ntot}$} &{\bf $\chi^2/{\rm ntot}$} & {\bf $\chi^2/{\rm ntot}$}  \\
& & & {\bf \#points, ntot} &$\Phi_{W-Y}$& $\Phi_{Cahn}$ & $\Phi_{Selyugin}$  \\
\hline
 & & & & & &\\
1& $\sigma_{total}^{\bar p p}$ & & 33  & 0.7167 & 0.7167 & 0.7167 \\
2& $\sigma_{total}^{p p}$ & & 68  &  0.3617 &  0.3617 &  0.3617 \\
3& $\rho^{\bar p p}$ & & 11  &  0.6086 &  0.6086 &  0.6086 \\
4& $\rho^{p p}$ & & 48  & 0.4326 & 0.4326 & 0.4326 \\
\hline
 & & & & & & \\
5& $\frac{d\sigma}{dt}^{\bar p p}$ & $24.3$ & 26  & 1.4965 & 2.4879 & 2.8105 \\
6& $\frac{d\sigma}{dt}^{\bar p p}$ & $30.4$ & 21  & 1.4702  &1.4130 & 1.4027 \\
7& $\frac{d\sigma}{dt}^{\bar p p}$ & $52.6$ & 12  & 1.9847  &1.4676 & 1.3440 \\
8& $\frac{d\sigma}{dt}^{\bar p p}$ & $62.3$ & 3  &  1.0263  &0.8135 & 0.7635 \\
9& $\frac{d\sigma}{dt}^{\bar p p}$ & $541.0$ & 37  &  2.0811 &2.3993 & 2.5246 \\
10& $\frac{d\sigma}{dt}^{\bar p p}$ & $546.0$ & 15  &  11.4692 &11.0364 &10.9008 \\
 & & & &  & &\\
\hline
 & & & &  & &\\
11& $\frac{d\sigma}{dt}^{p p}$ & $10.6$& 35  & 2.3342 &2.0297 & 1.9610\\
12& $\frac{d\sigma}{dt}^{p p}$ & $12.3$& 37  & 1.5731 &1.4169 & 1.3931 \\
13& $\frac{d\sigma}{dt}^{p p}$ & $19.4$& 45  & 1.9674 &1.7840 & 1.7509 \\
14& $\frac{d\sigma}{dt}^{p p}$ & $22.2$& 45  & 1.4666 &1.3374 & 1.3192 \\
15& $\frac{d\sigma}{dt}^{p p}$ & $23.9$ & 94  &  2.3910  &2.3699 &2.3087 \\
16& $\frac{d\sigma}{dt}^{p p}$ & $24.3$ & 25  & 1.1418  &1.2638 &1.3587  \\
17& $\frac{d\sigma}{dt}^{p p}$ & $27.4$ & 40  & 2.2246  &2.0014 &1.9607  \\
18& $\frac{d\sigma}{dt}^{p p}$ & $30.7$ & 8  &  0.3174 &1.2071 &1.5482   \\
19& $\frac{d\sigma}{dt}^{p p}$ & $44.7$ & 25  & 5.0147  &2.9964 &2.7679  \\
20& $\frac{d\sigma}{dt}^{p p}$ & $52.8$ & 21  & 7.1074  &4.2590 &3.9018  \\
21& $\frac{d\sigma}{dt}^{p p}$ & $62.3$ & 4  & 8.5177  &6.7054 & 6.2988  \\
\hline
\end{tabular}
\end{center}
\vskip -0.5cm
\caption{ $\chi^2$ per point for the region of low $|t|:\; 0\leq |t|\leq 0.01\:(GeV^2)$ for
$\Phi_{W-Y}$, $\Phi_{Cahn}$,
and $\Phi_{Selyugin}$. \label{tab:3}}
\end{table}
}

To appreciate the role of the Coulomb phase, we plot the angular 
distributions with the appropriate Coulomb phase and with the phase taken equal to zero in 
the following figures \ref{fig:ppdifct22},
\ref{fig:difpbarpct30}. 
Even though the difference is minimal, the numerical conclusion is that the data,
quite unambiguously, prefer the appropriate nonzero Coulomb phase.
 
The comparison with the West-Yennie and Cahn phases is not shown because the plot would be undistinguishable from the Selyugin one.

Let us remind that no additional fitting was done.

\begin{figure}[H]
{\vspace*{ -2cm} \epsfxsize=140mm \epsffile{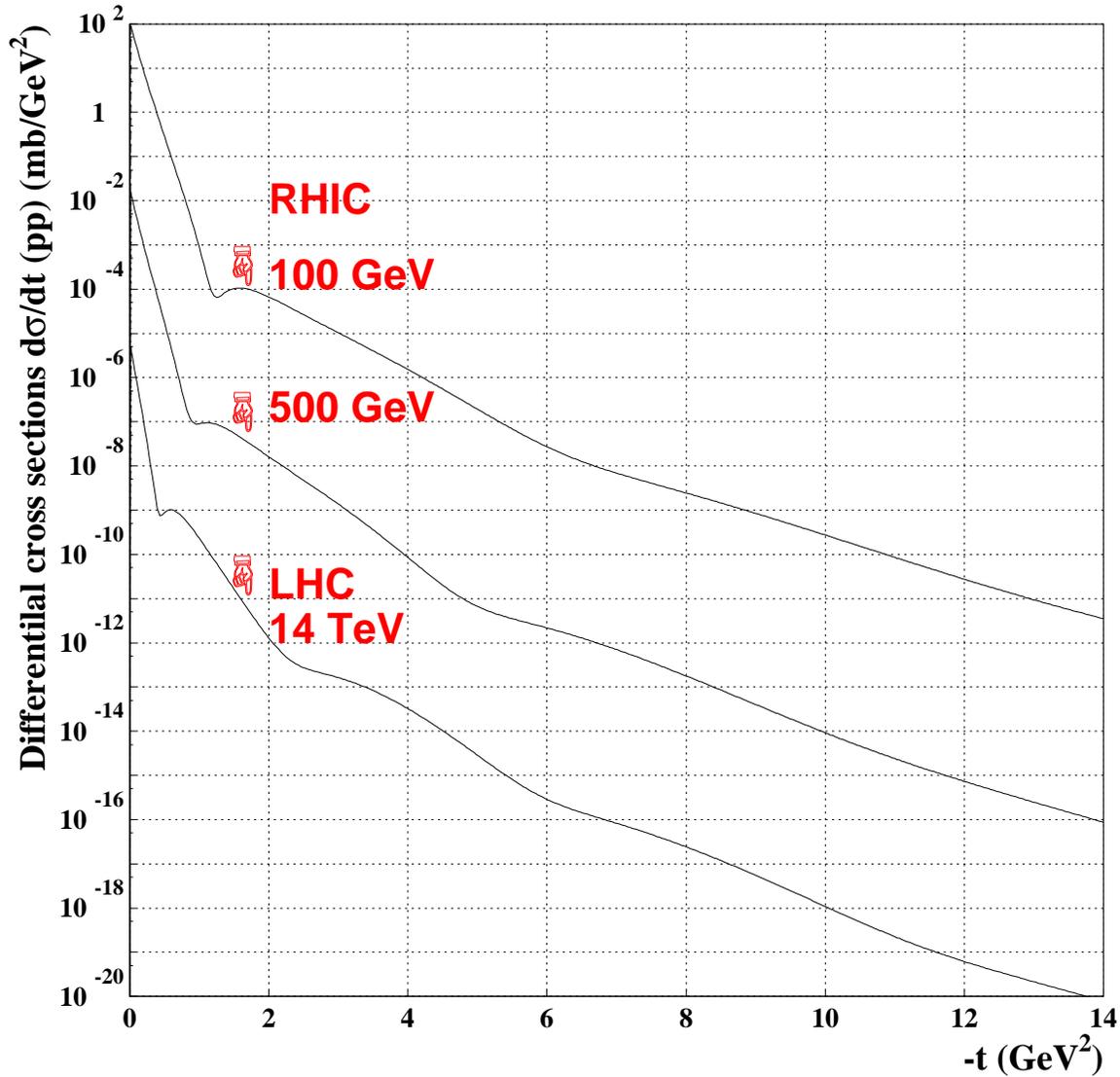}}
\vskip -3.cm
\caption{Prediction of the model for the $p p$ scattering differential cross section at {\bf RHIC} ($\sqrt{s}=100,\: 500\; GeV$)
and {\bf LHC} ($\sqrt{s}=14\; TeV$).
A $10^{-4}$ factor between each successive set of data is omitted.
The prediction for {\bf RHIC} at the energy of $\sqrt{s}=500\; GeV$ is multiplied by a factor of $10^{-4}$ and the prediction for {\bf LHC} at the energy of $\sqrt{s}=14\; TeV$ is multiplied by a factor of $10^{-8}$.
\label{fig:difpplhcc}
}
\end{figure}

\begin{figure}[H]
{\vspace*{ -2cm} \epsfxsize=140mm \epsffile{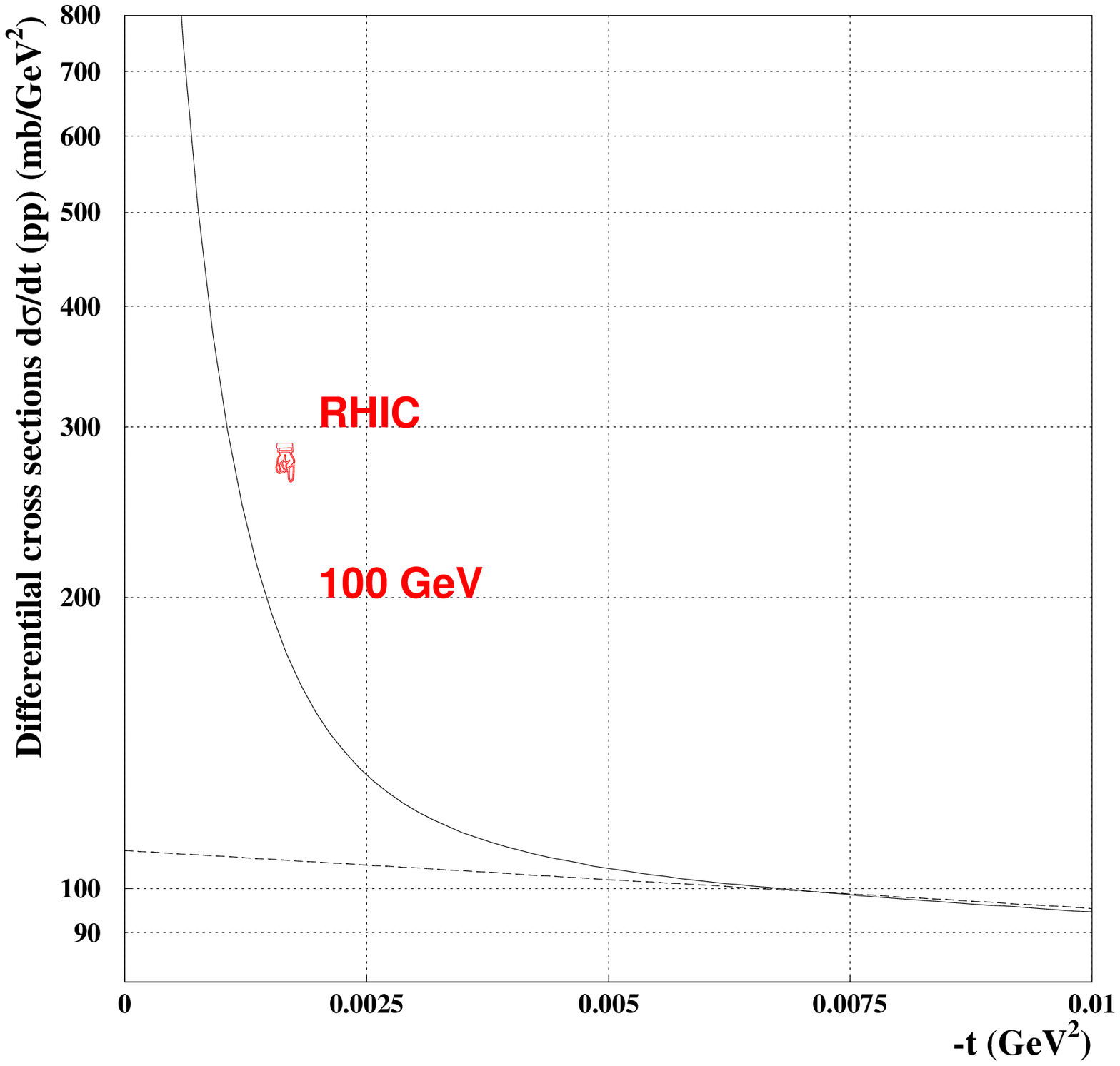}}
\vskip -3.cm
\caption{Prediction of the model for the $p p$ scattering differential cross section  at {\bf RHIC} ($\sqrt{s}=100\; GeV$).
The solid line
corresponds to full amplitude including the Coulombic one
and the dashed line corresponds to the nuclear amplitude \cite{threepomerons}.
\label{fig:difpprhic1t}
}
\end{figure}

\begin{figure}[H]
{\vspace*{ -2cm} \epsfxsize=140mm \epsffile{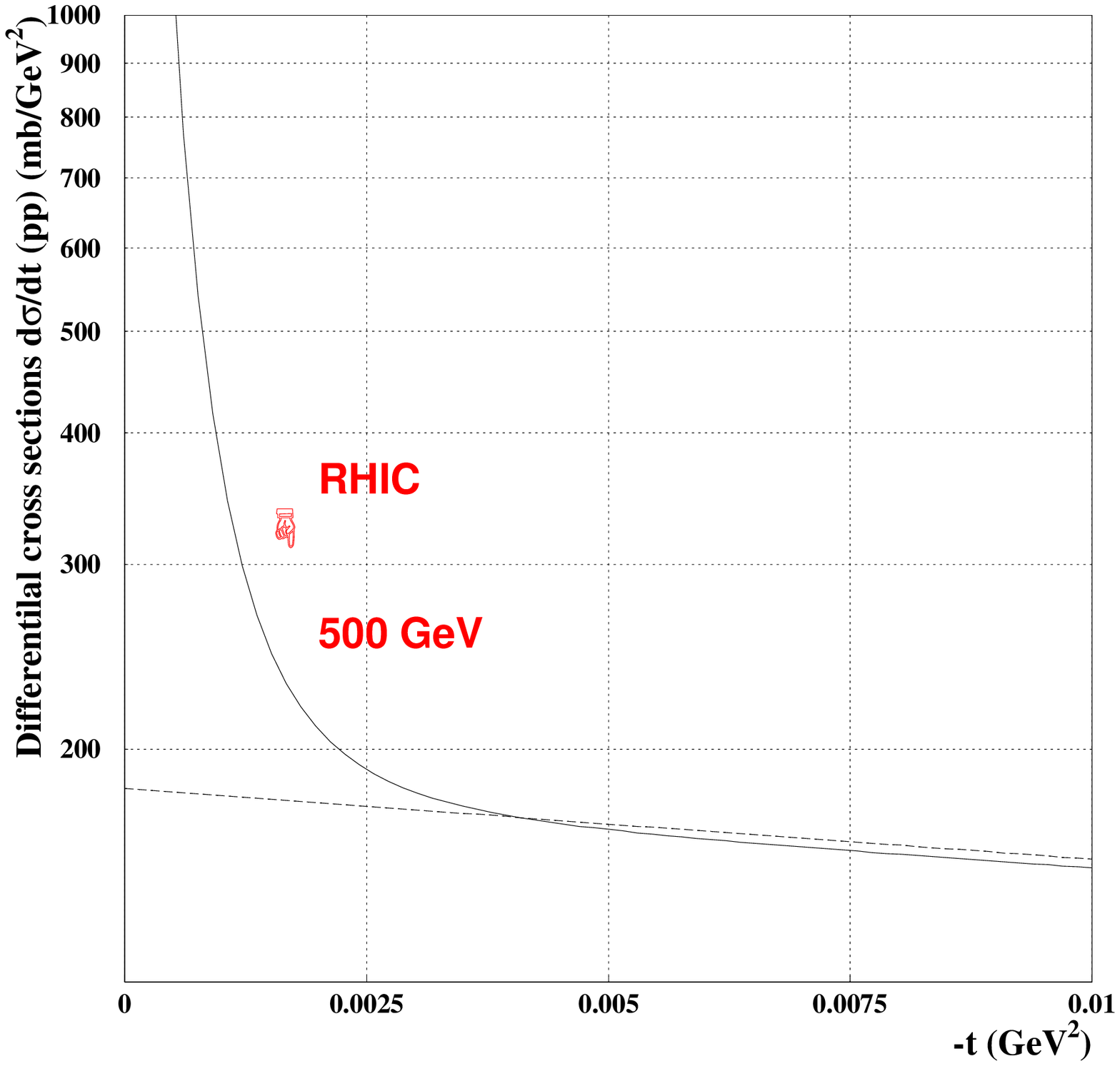}}
\vskip -3.cm
\caption{Prediction of the model for the $p p$ scattering differential cross section  at {\bf RHIC} ($\sqrt{s}=500\; GeV$).
The solid line
corresponds to full amplitude including the Coulombic one
and the dashed line corresponds to the nuclear amplitude \cite{threepomerons}.
\label{fig:difpprhic2t}
}
\end{figure}

\begin{figure}[H]
{\vspace*{ -2cm} \epsfxsize=140mm \epsffile{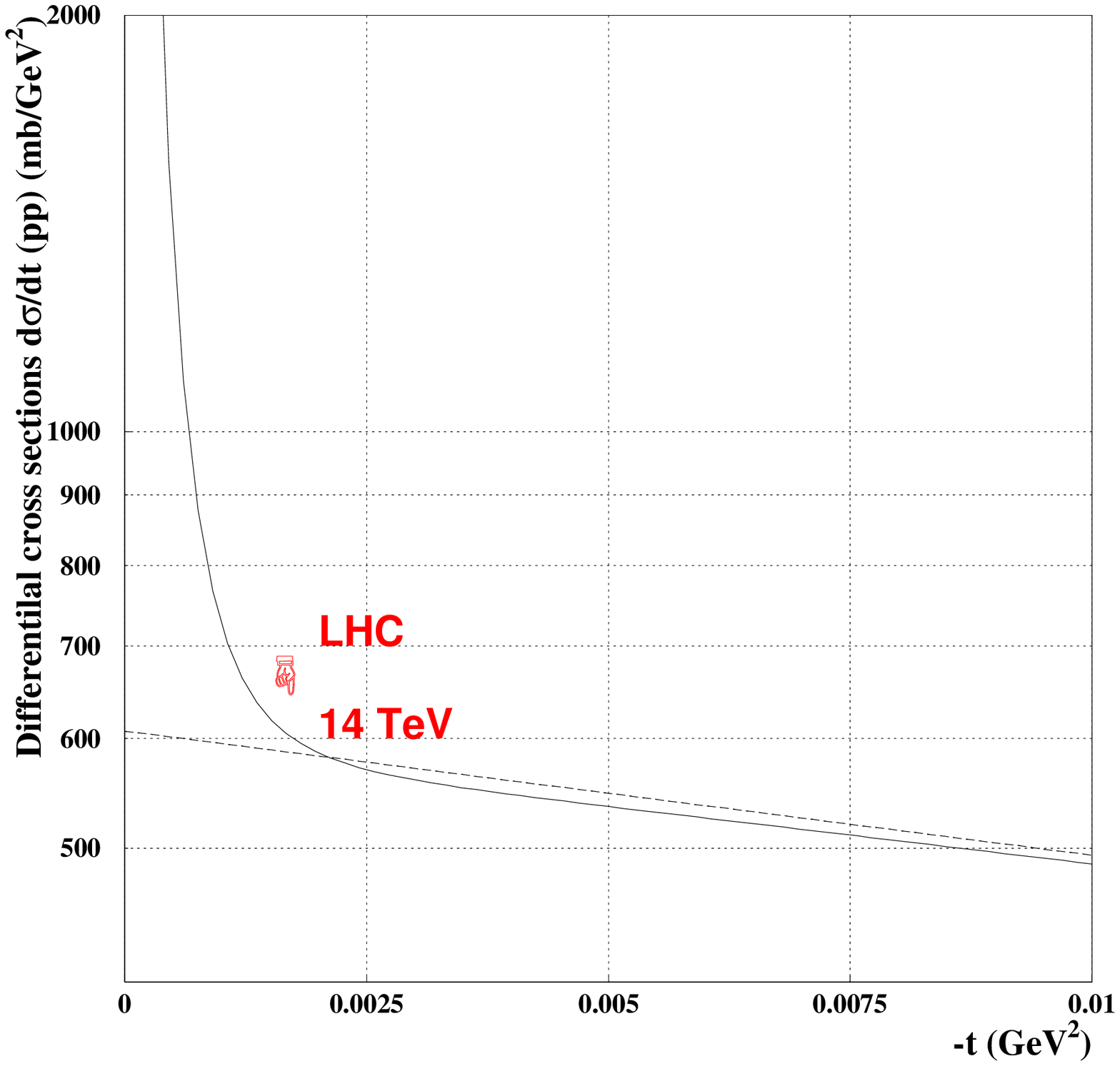}}
\vskip -3.cm
\caption{Prediction of the model for the $p p$ scattering  differential cross section  at  {\bf LHC} ($\sqrt{s}=14\; TeV$).  The solid line
corresponds to full amplitude including Coulombic one
and the dashed line corresponds to the nuclear amplitude \cite{threepomerons}.
\label{fig:difpplhcct}
}
\end{figure}

We do not report here any of the $pp$ and $\bar p p$ angular distributions which
one obtains over the full range of $|t|$ values because the original reproduction 
of these quantities is left basically unchanged by the Coulomb amplitude and the
interested reader is referred to the original paper~\cite{threepomerons}. We simply report here, for
completeness, the predictions at RHIC and LHC energies both in the 
interference region and over the entire $|t|$ range.  Predictions of the model 
and comparison with the nuclear
amplitude for {\bf RHIC} and {\bf LHC} are shown in figures
 \ref{fig:difpplhcc}, \ref{fig:difpprhic1t}, \ref{fig:difpprhic2t}, \ref{fig:difpplhcct}.

\begin{figure}[H]
\vskip -0.5cm
\parbox[c]{6.5cm}{\vspace*{ -0.5cm} \epsfxsize=65mm \epsffile{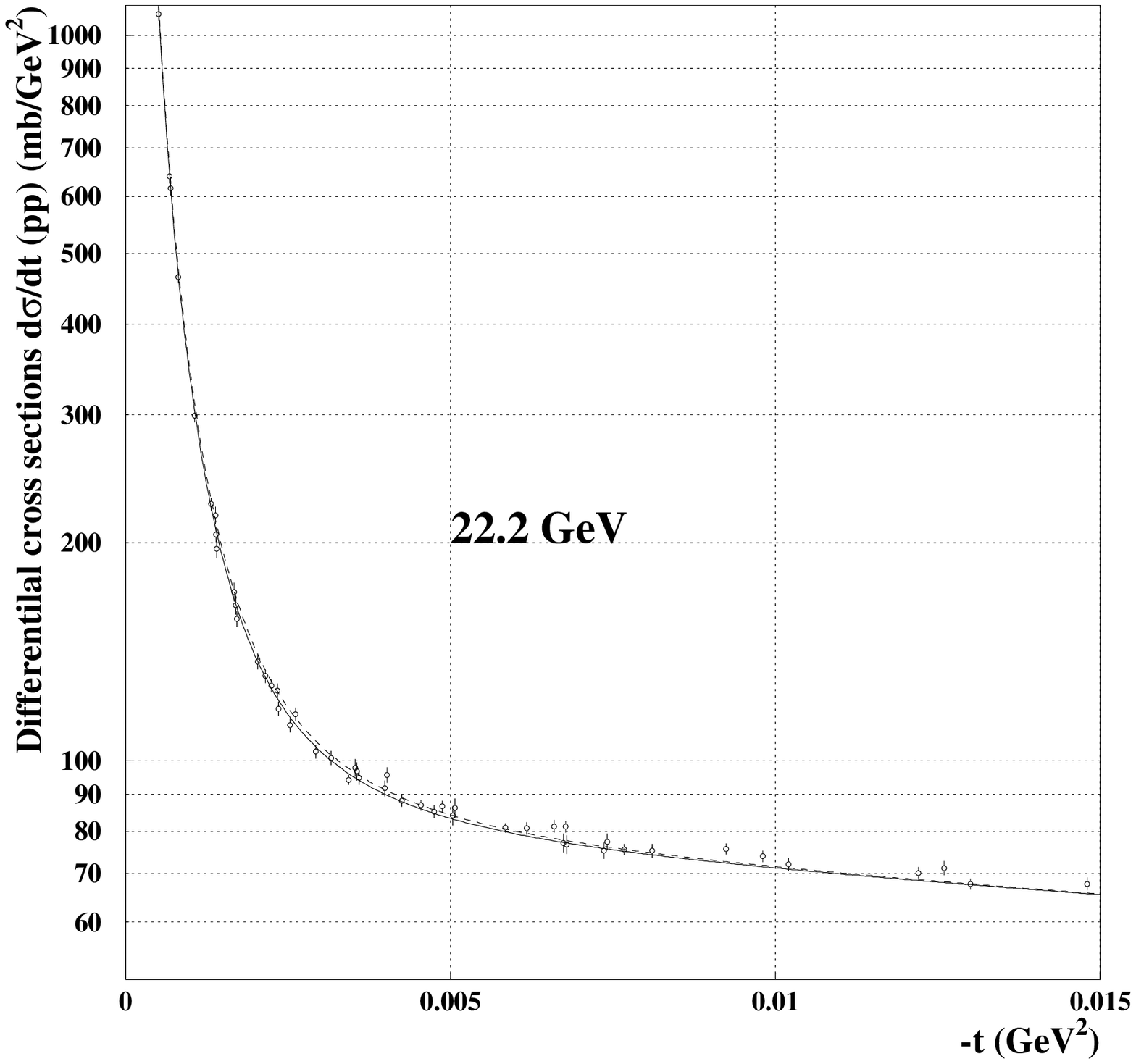}}
\hfill~\parbox[c]{6.5cm}{\vspace*{ -0.5cm} \epsfxsize=65mm 
\epsffile{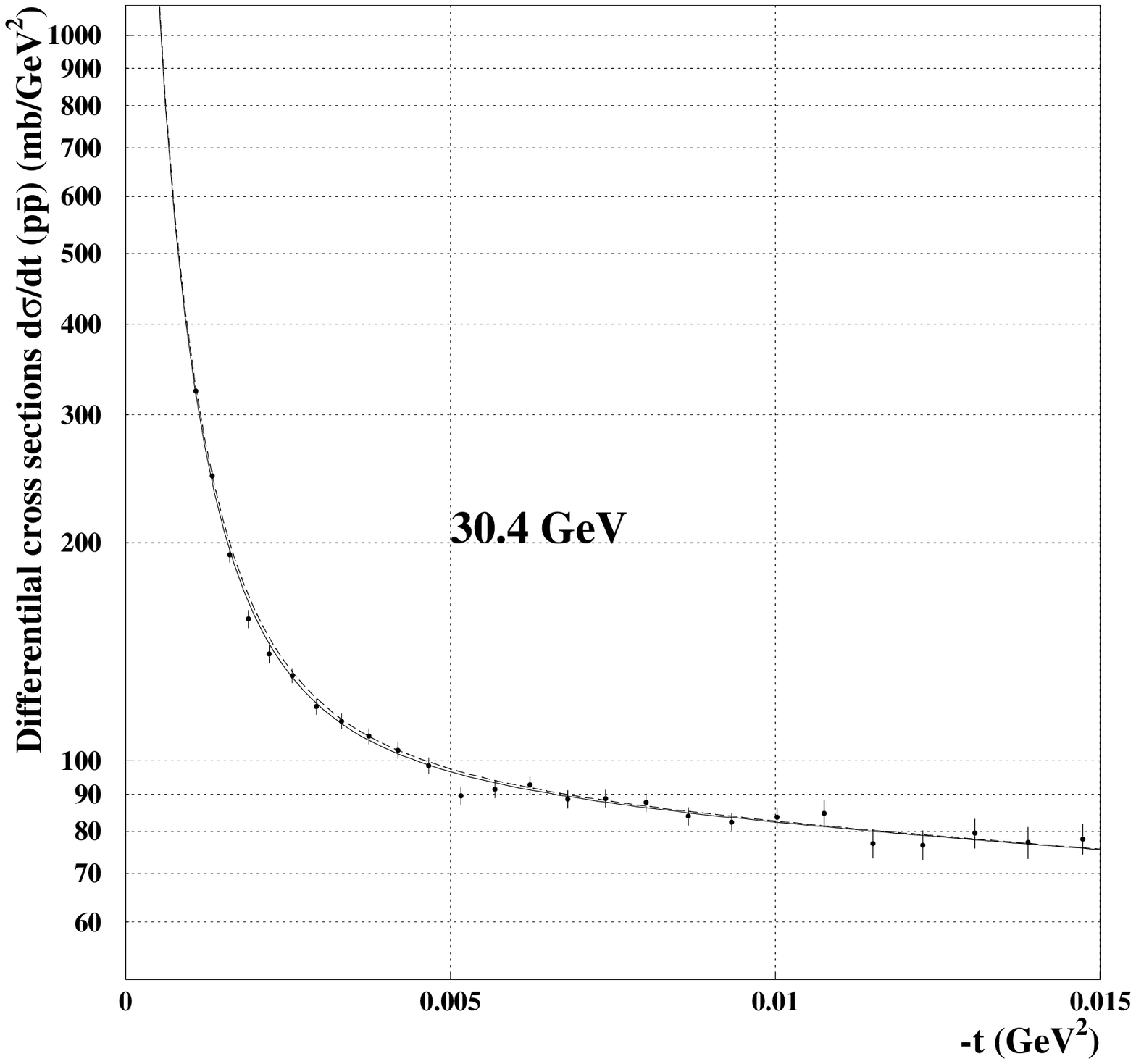}}

\vskip -1.5cm

\parbox[t]{6.5cm}{\footnotesize 
\caption{Differential cross section of $p p$ scattering
and curves corresponding to its description in the model. The solid line
corresponds to the Coulomb phase calculated with prescription of Selyugin
(\ref{eq:selphase}) and the dashed line corresponds to the zero phase.
\label{fig:ppdifct22}
}
}
\hfill~\parbox[t]{6.5cm}{\footnotesize 
\caption{Differential cross section of $\bar p p$ scattering
and curves corresponding to its description in the model. The solid line
corresponds to the Coulomb phase calculated with prescription of Selyugin
(\ref{eq:selphase}) and the dashed line corresponds to the zero phase.
\label{fig:difpbarpct30}
}
}

\end{figure}


\begin{figure}[H]
\vskip -0.5cm
\parbox[c]{6.5cm}{\vspace*{ -0.5cm} \epsfxsize=65mm \epsffile{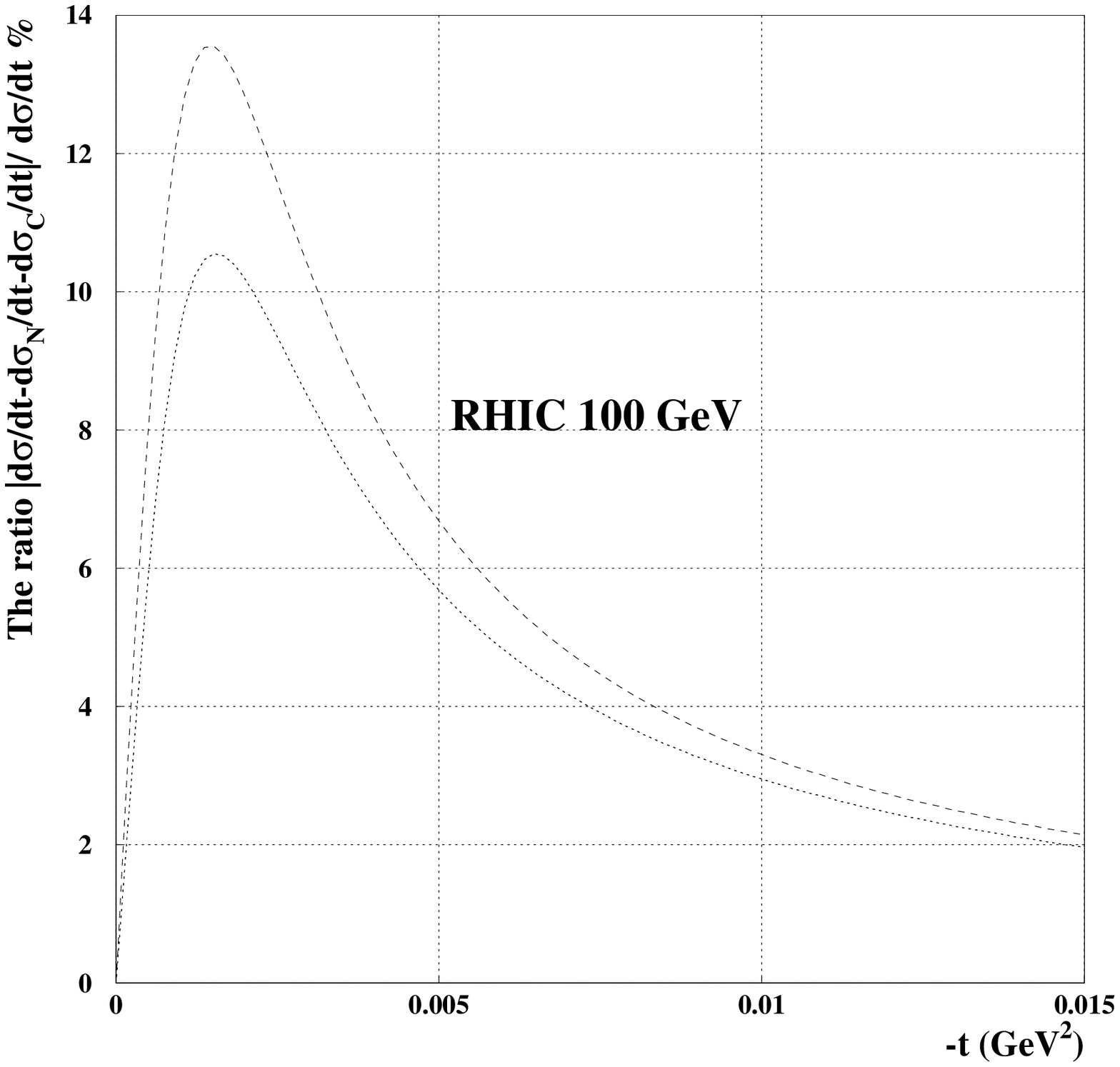}}
\hfill~\parbox[c]{6.5cm}{\vspace*{ -0.5cm} \epsfxsize=65mm 
\epsffile{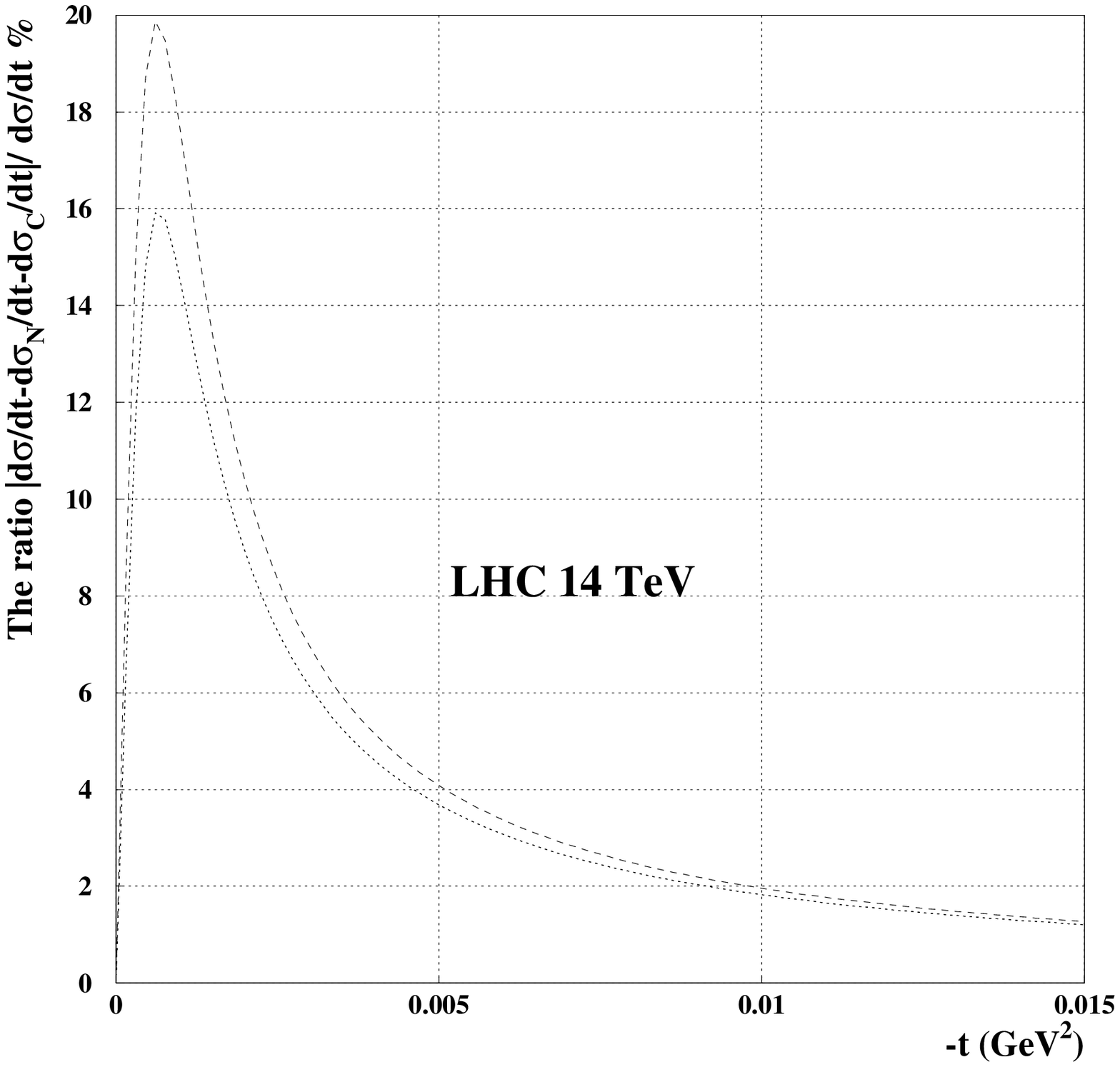}}

\vskip -1.5cm

\parbox[t]{6.5cm}{\footnotesize 
\caption{The interference term between hadron and Coulomb amplitudes
for at the  {\bf RHIC} energy.
The dashed line corresponds to the Coulomb phase calculated with the prescription of Selyugin \cite{Selyugin} and the dotted line to the zero Coulomb phase.
\label{fig:interferencerhic}
}
}
\hfill~\parbox[t]{6.5cm}{\footnotesize 
\caption{The interference term between hadron and Coulomb amplitudes
at the {\bf LHC} energy.
The dashed line corresponds to the Coulomb phase calculated with the 
prescription of Selyugin \cite{Selyugin} and the dotted line to the zero Coulomb phase.
\label{fig:interferencelhc}
}
}
\end{figure}

To show the influence of the interference between nucleon and
Coulomb amplitudes we show the ratio
\be
R = \frac{\Big| \frac{d\sigma}{dt}-\frac{d\sigma_{nucleon}}{dt}-\frac{d\sigma_{coulomb}}{dt} \Big| }{\frac{d\sigma}{dt}}, \;[\%]
\ee
in figures \ref{fig:interferencerhic} and \ref{fig:interferencelhc} 
and we conlude that the interference term becomes negligible at 
{\bf RHIC} and {\bf LHC} energies at $t\sim -0.01\; GeV^2$ for {\bf RHIC}
and $t\sim -0.005\; GeV^2$ for {\bf LHC}. 

Now let us return to the problem of extracting the ratio $\rho(s,t=0)$ 
of the
real to the imaginary parts of the scattering amplitude. A general theorem
\cite{Martin} shows that this ratio as a function of the momentum transfer
could not have a constant sign in a strip $s_M < s < \infty$, $-T<t\le 0$,
for any $s_M$ and $T>0$. The method of extracting the ratio includes a
 simplification, the scattering amplitude is usually presented
in the following form:

\be
\displaystyle T(s,t)=\mp \frac{8\pi \alpha s}{|t|}f^2(|t|)
+(i+\rho(s,0))s\sigma_{tot}e^{Bt}(1 - i\alpha\Phi)
\ee
where the ratio $\rho(s,t)$ is approximated by its value at $t=0$. 
Should this simplification affect the measurement?
The answer would be ``yes'' should the change of sign of $\rho$ occurr
close to the region of the maximum of the Coulomb nuclear interference. 

\begin{figure}[H]
\vskip -0.5cm
\parbox[c]{6.5cm}{\vspace*{ -0.5cm} \epsfxsize=65mm \epsffile{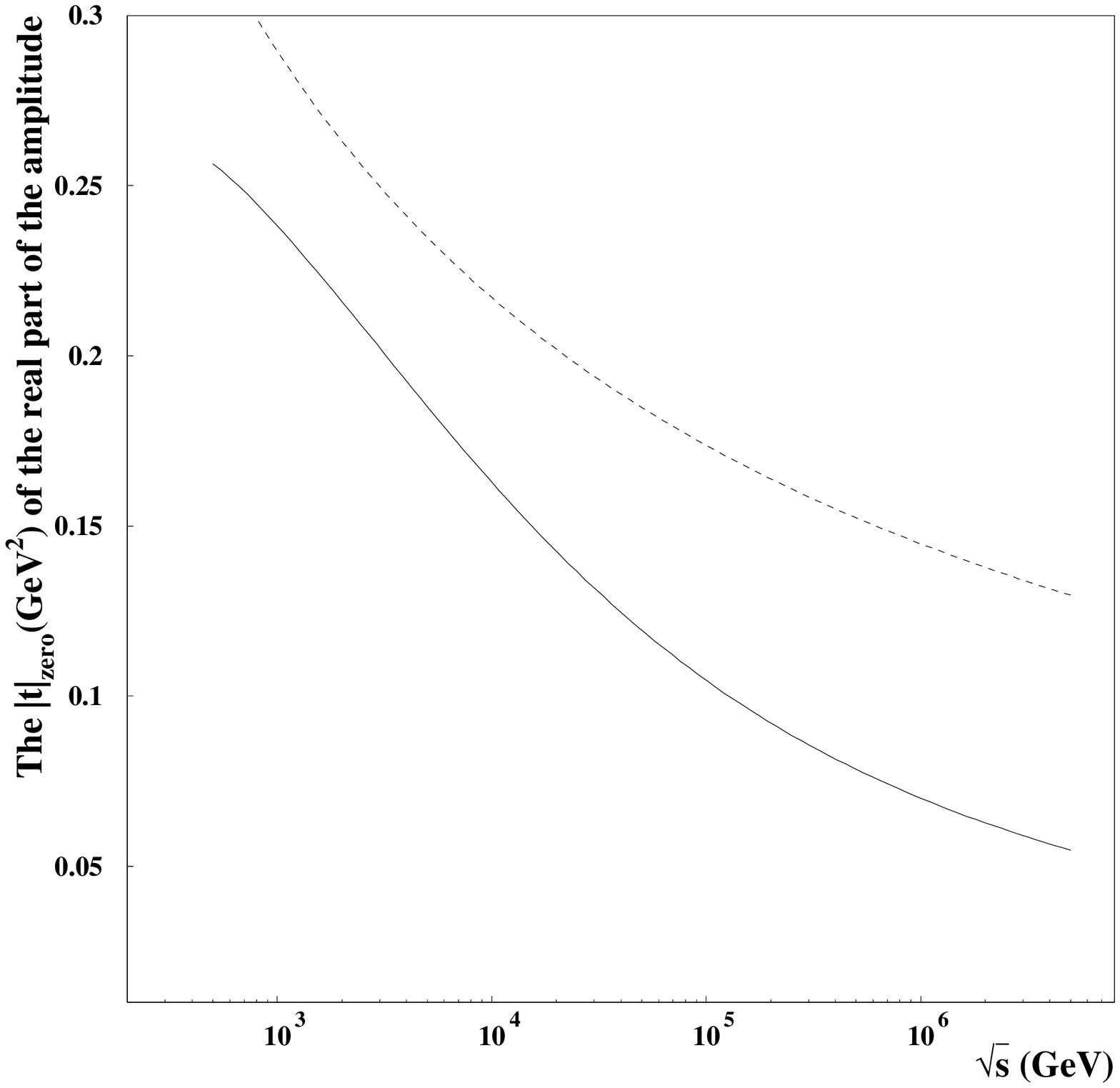}}
\hfill~\parbox[c]{6.5cm}{\vspace*{ -0.5cm} \epsfxsize=65mm 
\epsffile{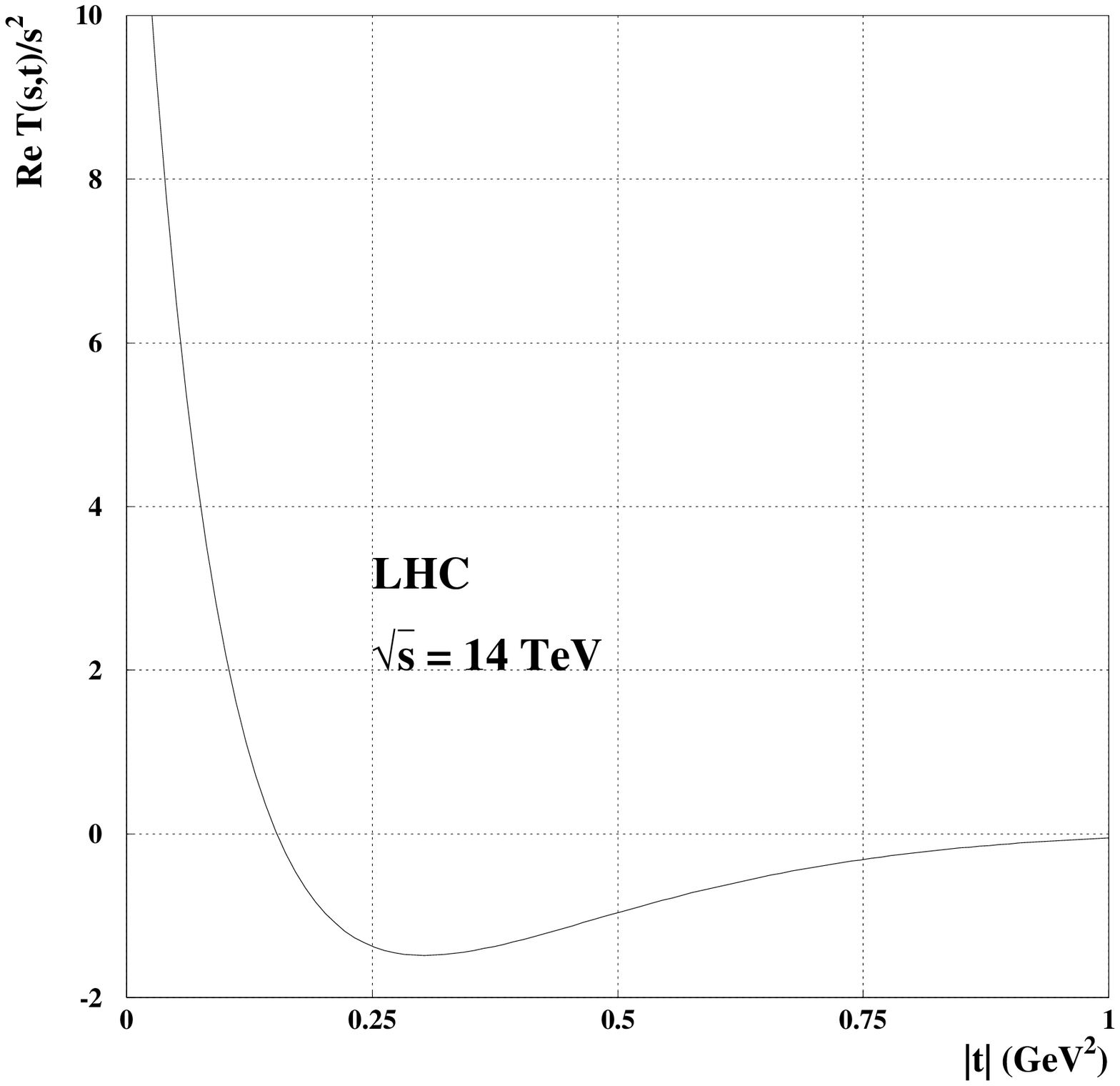}}

\vskip -1.5cm

\parbox[t]{6.5cm}{\footnotesize 
\caption{The position of the zero of the real part of the scattering amplitude
in a three-component Pomeron model \cite{threepomerons} (solid line) and
in a ``geometrical scaling'' model $t_{zero}=-1/(\lambda \ln s)$, $\lambda =
0.25 \; GeV^{-2}$ \cite{Martin} (dashed line).
\label{fig:zero}
}
}
\hfill~\parbox[t]{6.5cm}{\footnotesize 
\caption{The real part of the scattering amplitude
at the {\bf LHC} energy in the  model \cite{threepomerons}.
\label{fig:zeroLHC}
}
}
\end{figure}

In our particular model, the zero of the real part of the scattering amplitude
goes to zero with the increase of energy, but remains far from the
region of the maximum of the interference. For example, at LHC energies
$\sqrt{s}=14\; TeV$, the maximum occurs at $|t_{max}|\sim 10^{-3}\;GeV^2 $ (fig.\ref{fig:interferencelhc}) while
the zero is at $|t_{zero}|\sim 0.16\;GeV^2$ ( fig. \ref{fig:zero}, \ref{fig:zeroLHC}), so the situation is not too bad. 
\section*{CONCLUSION}

All the three choices of the Coulomb phase
give good description of the existing data, though in terms of $\chi^2$ per point the data ``prefer'' the phase calculated with the prescriptions of 
Cahn \cite{Cahn} (\ref{eq:ourcahn}) and Selyugin \cite{Selyugin} (\ref{eq:selphase}). We conclude that such prescriptions are a
good basis for the evaluation of the Coulombic contribution in the full
scattering amplitude.  We expect that similar conclusion would apply to an
analysis repeated along the lines of \cite{Kop} but we have not performed this analysis.

As we have seen, the addition of the nuclear amplitude (with parameters fitted 
from total and differential cross sections) and of the Coulomb one (with its 
proper phase) gives a total amplitude which reproduces quite well the data in 
the interference region without any additional parameters and with no need to
refit existing ones. 

This allows us to predict the {\bf RHIC} Coulomb interference which requires the
measurements to start from $|t|\le0.005\;GeV^2$ at the energy of $\sqrt{s}=100\; GeV$ 
and from $|t|\le 0.004\;GeV^2$ at the energy of $\sqrt{s}=500\; GeV$.
Likewise, {\bf LHC} will be able to cover the Coulomb region if the measurement starts 
from $|t|\le 0.001\;GeV^2$.

\section*{ACKNOWLEDGMENTS}
We would like to thank E. Martynov and O. Selyugin for useful discussions.


\end{document}